\documentclass[12pt,preprint]{aastex}
\usepackage{epsfig}
\usepackage{psfig}

\begin{document}

\title{Observing Conditions at Mt.~Graham: VATT \boldmath $UBVR$ Sky 
Surface Brightness and Seeing Measurements from 1999 through 2003.}

\author{Violet A. Taylor\email{violet.taylor@asu.edu}}
\author{Rolf A. Jansen\email{rolf.jansen@asu.edu}} 
\and 
\author{Rogier A. Windhorst\email{rogier.windhorst@asu.edu}}
\affil{Department of Physics and Astronomy, Arizona State University,
    Box 871504, Tempe, AZ 85287-1504}

\begin{abstract}

We present measurements of sky surface brightness and seeing on
Mt.~Graham obtained at the Vatican Advanced Technology Telescope (VATT)
during 16 observing runs between April 1999 and December 2003.  We show
that the sky surface brightness is significantly darker during
photometric conditions, and can be highly variable over the course of a
single observing run as well as from one run to the next, regardless of
photometricity.  In our photometric observations we find an average
low-airmass ($\sec z < 1.2$) sky surface brightness of 22.00, 22.53,
21.49, and 20.88 mag\,arcsec$^{-2}$ in $U$, $B$, $V$, and $R$,
respectively.  The darkest run (02/00 in $U$ and 02/01 in $BVR$) had an
average sky surface brightness of 22.38, 22.86, 21.72, and 21.19
mag\,arcsec$^{-2}$ in $U$, $B$, $V$, and $R$, respectively.  With these
results we show that under the best conditions, Mt.~Graham can compete
with the darkest sites in Hawaii and Chile, thanks in part to the strict
dark-sky ordinances in place in Tucson and Safford.  We expect the sky
over Mt.~Graham to be even darker than our 1999--2003 results during
solar minimum (2006--2007). 

We find a significant improvement of about $0.45\arcsec$ in our measured
stellar FWHM after improvements to the telescope were made in Summer and
Fall 2001.  Stellar FWHM values are highly variable, with median
$R$-band focus FWHM values in each observing run ranging from
0.97\arcsec\ to 2.15\arcsec.  Significantly sub-arcsecond seeing was
occasionally achieved with values as low as $0.65\arcsec$ FWHM in $R$. 
There may possibly still be a significant telescope contribution to 
the seeing at the VATT, but nearby trees as high as the dome are currently the
dominant factor. 

\end{abstract}


\section{Introduction}

Mount Graham International Observatory (MGIO) is located near Safford, 
Arizona at an altitude of 10,400 feet. It contains the Vatican Advanced 
Technology Telescope (VATT), the Heinrich Hertz Submillimeter Telescope, 
and the Large Binocular 
Telescope\footnote{\url{http://medusa.as.arizona.edu/lbto/}} 
(LBT; currently under construction with first light expected in late 2004). 
The observing conditions
at the MGIO site are important limiting factors on the efficiency
of observing faint objects, and are thus important to characterize with
observations at the existing telescopes, as well as the LBT. Therefore,
in this paper we focus on two of the most important properties of
an observing site: the sky surface brightness and the seeing over the
course of four years.

Dark sites are in increasingly short supply due
to metropolitan development, but reasonably dark sites do still exist.
Other observers have studied sky surface brightness values at other
observing sites, particularly in the context of determining the effects
of nearby city lights. Massey and Foltz (2000) measured the sky
brightness in various directions of the sky at Kitt Peak and Mt.~Hopkins
in 1988 and again in 1998 to determine the effects of increasing light 
pollution from the expansion of Tucson. They found that since 1988, the zenith 
$BV$ sky brightness increased slightly by 0.1--0.2 mag\,arcsec$^{-2}$ at
Kitt Peak. At a larger zenith distance of $60\degr$, however, there
was a 0.35 mag\,arcsec$^{-2}$ increase when pointing away from Tucson,
and a 0.5 mag\,arcsec$^{-2}$ increase when pointing toward Tucson. They 
mention that this increase in sky brightness would be worse if Tucson
did not have good outdoor lighting ordinances, which also exist in
Safford. Although Mt.~Graham is near Safford, Safford is a much smaller 
city than Tucson and MGIO is located at a much higher elevation than 
Kitt Peak and Mt.~Hopkins, with Tucson and Phoenix well below the horizon
as viewed from the Mt. Graham summit. Hence, city lights should not 
have as large of an impact on the sky brightness at MGIO.

Other factors in addition to city lights impact the sky brightness, such
as the presence of atmospheric dust, forest-fire smoke, cirrus, the 
solar cycle, airmass, 
galactic and ecliptic latitude of the observation, the phase and angular 
distance of the Moon from the observed object, and altitude and geomagnetic 
latitude of the 
observing site. Benn and Ellison (1998) measured the sky brightness at La Palma
from 1987 to 1996, finding that the sky was 0.4 mag\,arcsec$^{-2}$ brighter 
during solar maximum than solar minimum, and 0.25 mag\,arcsec$^{-2}$ brighter 
at an airmass ($\sec z$) of 1.5 than an airmass of 1.0 (at the zenith).
Krisciunas (1997) measured the sky brightness at Mauna Kea in Hawaii,
and found that except for the solar cycle, the most important effect
is random short term variations over tens of minutes, which makes
sky brightness measurements highly variable and difficult to compare
between sites. To quantify the quality of sky brightness at Mt.~Graham, 
we present our sky surface brightness measurements from 
April 1999 to April 2002 at the VATT, compare our measurements to those 
known at 
Mt.~Hopkins, Kitt Peak, Mauna Kea, La Palma, ESO, and Cerro Tololo, and 
discuss how the variability of sky brightness due to the factors listed 
above impact our conclusions. We also compare our measurements to a
theoretical sky brightness for Mt.~Graham (Garstang 1989) and investigate
the effects of city lights and the variation of sky brightness with time
of night.

The seeing of an astronomical site can be estimated by measuring the 
median full width at half max (FWHM) of stars in images taken at that site.
We have done this for Mt.~Graham by measuring the FWHM of stars in 
stacked galaxy images and in short focus exposures taken at the VATT. This is 
only an estimate, because there are other factors in addition to 
atmospheric seeing that play a role in the stellar FWHM, such as 
telescope focus and telescope image quality due to mirror quality,
telescope collimation, etc.. The FWHM results presented 
in this paper are to be applied at face value to 
the VATT alone, and do not necessarily reflect on the Mt.~Graham site or on 
the LBT site, since the VATT's specific location on the mountain-top makes it 
more susceptible to ground layer seeing, particularly in northeasterly winds.

\section{Observations}

We have obtained $UBVR$ surface photometry for 142 galaxies
at the VATT, using the VATT 2k$\times$2k Direct CCD Imager.
Typical exposure times were
2$\times$(600--1200)s in $U$, 2$\times$(300--600)s in $B$, 
2$\times$(240--480)s in $V$, and 2$\times$(180--360)s in $R$. 
The CCD gain is 1.9 electrons per ADU and the read-noise is 
5.7 electrons. We binned the images 2$\times$2, resulting in a pixel
scale of 0.375 arcsec pixel$^{-1}$. Individual images were stacked with
integer shifts, as the PSF is well sampled. Sky brightness values and FWHM 
values measured from
stacked images are the signal-to-noise weighted average values from the 
individual images that make up the stack, which suffices to examine overall
trends in the data. The details of our galaxy sample and galaxy surface
photometry, and the methods 
we used for data reduction and calibration are presented in a separate
data paper on our nearby galaxy survey (V. Taylor, et al., in preparation). 

Observations were spread over 9 runs between April 1999 and April 2002,
for a total of 49 usable nights. Defining photometric nights as those
with zeropoint magnitudes that vary no more than 3\% throughout, 
45\% of the nights were photometric, 51\% were mostly non-photometric
(with parts of the night possibly photometric until clouds moved in), 
and 4\% were lost entirely 
to telescope problems. During nights where clouds appeared toward the
end of the night, we salvaged as much as possible of the first part of 
the night as photometric. 

For comparison, additional focus exposure stellar FWHM values are  
presented for 8 VATT observing runs between November 2001 and December 2003, 
which were carried out independently by R.A. Jansen for other projects. 

\section{Trends in Sky Surface Brightness at the VATT}

\subsection{Measurements of the Sky}

Sky values for each stacked galaxy image were calculated
by finding the median of the median pixel value in each of 13 boxes,
each 120 pixels wide, along the edges of the image. This was done
to avoid including light from the galaxy, which was usually centered in the
CCD. Taking the median values helps to reject stars and cosmic
rays, which comprise a small percentage of the total number of
pixels in the sky boxes. The
average sky count-rates for all stacked galaxy images were $0.41 \pm 0.01$ 
ADU s$^{-1}$ in $U$, $1.34 \pm 0.11$ ADU s$^{-1}$ in $B$, 
$2.64 \pm 0.10$ ADU s$^{-1}$ in $V$, and
$4.26 \pm 0.15$ ADU s$^{-1}$ in $R$. Sky surface brightness values 
were photometrically calibrated using Landolt standards (Landolt 1992). 
We defined photometric 
nights as those with zeropoints that vary no more than 3\% throughout the 
night, which defines the largest uncertainty in the calibrations. 

\subsection{Sky Surface Brightness Results}

In Figure 1, the resulting $UBVR$ sky surface brightness values for each stacked
galaxy image are plotted vs. the average airmass ($\sec z$) of the 
individual images that comprise each stack. Each observing run is broken up
into a separate panel for comparison. Stacked images that are
comprised solely of individual images taken during photometric conditions
(change in magnitude zeropoint throughout the night $\la 3\%$) are plotted as 
asterisks, while those comprised of images taken during non-photometric
conditions are plotted as open circles. There is a clear, well defined 
difference in
sky surface brightness between these two conditions: non-photometric
nights have notably brighter skies, as expected due to 
the presence of cirrus. There is a trend of increasing
sky surface brightness with increasing airmass, which is also to
be expected, although there does not appear to be a single consistent
slope to this trend throughout all observing runs, even 
for photometric runs. It is also apparent from the plots in Figure 1
that the sky surface brightness is highly variable as a function
of time, both over the course of a single run and from one run to the
next. Since the sky brightness is highly dependent on many factors,
such as solar activity, atmospheric conditions, time since sunset,
variable night sky-lines, and the location of
the telescope pointing with respect to nearby city lights, the Moon,
zodiacal light, and the Galaxy itself, this variability is not
surprising. 

The effect of the Moon on the sky surface brightness of a given galaxy
field is a complicated function of the phase of the Moon, the airmass
($\sec z$) of both the Moon and the galaxy position, the angular distance 
between the Moon and the galaxy ($\theta_{Mg}$), and the atmospheric 
extinction (Krisciunas \& Schaefer 1991). 
We approximate the effect of Moon on our sky surface brightness 
($\mu$) results through a plot of the sky brightness of 
all stacked galaxy images vs. $\cos\theta_{Mg}$, which is 
shown in Figure 2. We use $\cos\theta_{Mg}$
because the effects from the Moon on the sky brightness of a target away 
from the Moon may behave as a spherical harmonic, so that some linear 
behavior in $\cos\theta_{Mg}$ may be expected.
The secondary Moon effects due to airmass of the Moon 
and galaxy and atmospheric extinction are not separated out here, and are
expected to be small compared to other large scale variations in the
overall sky surface brightness, as discussed previously. The sky brightness
values were normalized to the median sky brightness for the relevant observing
run in order to remove large-scale seasonal effects. Four sub-panels
show different Moon phases, ranging from a Moon illumination of 0\%
to 40\%, which is the maximum illumination in our
data. This plot shows that photometric nights (indicated
by solid symbols) tend to have darker skies than non-photometric
nights (open symbols), and show a smaller scatter in
sky surface brightness from one image to the next. Both photometric
and non-photometric exposures show no major trend with Moon angular distance 
within the scatter for Moon illumination $\le 20\%$. There may
be a slight anti-trend of increasing sky surface brightness at $180\degr$
from the Moon visible in the panel for Moon illumination $\le 10\%$, 
which could be the
result of sun-light back-scattering off of the atmosphere. A stronger
trend of increasing sky surface brightness with decreasing Moon angular 
distance is apparent when Moon illumination is $\ge 20\%$. We applied a
linear least-squares fit of
\begin{equation}
\mu = m\,\cos\theta_{Mg} + b
\end{equation}
to this trend for the mostly photometric data in the 
Moon illumination $\ge 30\%$ panel (eliminating points with airmass
$> 2$), and determined slopes of
0.97 in $U$, 0.83 in $B$, 0.36 in $V$, and $0.29$ in $R$. Thus, as  
expected, there is a stronger dependence on Moon angular distance
for shorter wavelengths. There are only a small number ($\sim$5) of galaxy 
images that are affected by the Moon within the scatter of
these plots, leading us to the conclusion that our median sky
surface brightness values are largely unaffected by moonlight. 

Solar maximum occurred
around 2000--2001, in the middle of the time spanned by our observations, 
which could
have raised the sky surface brightness by several tenths of a magnitude
with respect to the sky surface brightness at solar minimum. For instance, 
Benn and Ellison (1998) saw an increase
in sky brightness of 0.4 magnitudes in $UBVR$ from solar minimum to solar 
maximum at La Palma. We therefore expect the sky surface brightness
to be fainter than these results by a similar amount during the upcoming 
solar minimum (2006--2007). 

The dotted lines in Figure 1 ($B$ and $V$ panels) represent an estimate 
of the dependence of the sky surface brightness on airmass at 
Kitt Peak and Mt.~Hopkins as measured by Massey and Foltz (2000) 
for comparison. Zenith values (airmass=1.00) were
derived by taking the average of Massey and Foltz's measurements at both
locations, which consisted of 1 exposure in each passband at
Mt.~Hopkins in Nov. 1998, and 4 exposures in each passband at
Kitt Peak over three nights in Nov. 1999 (all of which were just before 
solar maximum, like our earlier runs. However, our later runs are
closer to the solar maximum peak, and thus will be brighter). We calculated 
an average high airmass sky surface brightness by taking the average of 4
exposures in each passband at Mt.~Hopkins at zenith distances of $34-53
\degr$, and 6 exposures in each passband at Kitt Peak at zenith distances
of $\sim$60\degr. The dotted lines in Figure 1 connect these
two points, assuming a linear dependence on airmass, which is roughly correct.
One should be cautious when comparing sky brightness
measurements for different sites, due to the strong variability over
time visible in these figures, especially in a case like Mt.~Hopkins/Kitt
Peak, where we have no information on long-term variations. An additional
source of uncertainty arises because
Massey and Foltz (2000) derived their broadband sky brightness values from
spectrophotometry, replacing the variable \ion{O}{1} $\lambda$5577 line with
an average value. Nonetheless, we can see
that several VATT observing runs had sky surface
brightness values that were significantly darker than the Mt.~Hopkins/Kitt
Peak numbers given by Massey and Foltz (2000), who point out that their numbers
are comparable to Palomer Observatory in the early 1970's, which was
considered a rather dark site at the time.

Figure 3 shows the median sky surface brightness at the VATT, per observing
run, of all
low airmass ($\sec z < 1.2$) stacked galaxy images taken during photometric
conditions, as a function of time. One obvious outlying
sky brightness value was rejected in the 1/02 run in $U$ and $B$, which
was measured near morning twilight, and therefore contaminated our results.
For comparison, we overlay the average values from Mt.~Hopkins/Kitt Peak, 
(Massey \& Foltz 2000), Cerro Tololo, (Walker 1987, and 
1987--1988\footnote{\url{http://www.ctio.noao.edu/site/pachon\_sky/}}), 
and La Palma, (Benn \& Ellison 1998). Again, we caution against
putting strong confidence in such comparisons for the reasons previously
mentioned. On occasion the VATT was darker than Cerro Tololo, except
in the $V$-band. The La Palma observations were taken from 1987 to 1996, and
the values plotted in Figure 3 are the solar minimum values given by
Benn and Ellison (1998) minus the quoted 0.4 magnitude difference between 
solar minimum
and maximum, since our data was taken near solar maximum. For the
most part, our values are consistently darker than La Palma's solar maximum
skies, and similar to La Palma solar minimum skies (sometimes
brighter, sometimes darker, although always brighter in the $V$-band).
Figure 3 clearly shows a strong variability of several tenths of a
magnitude in sky brightness from observing run to observing run,
with a general brightening of the sky toward 
solar maximum (2000--2002). The anomalously bright
point during solar minimum in 1999 may have been due to smoke from
nearby forest fires.

In Table 1 we list our average photometric low 
airmass sky surface brightness values for the VATT, Mt. Graham, and
for various other sites for comparison. We also give sky surface 
brightness values for our darkest and brightest runs.
Excluding the Mauna Kea solar maximum values (Krisciunas 1997), which 
are significantly brighter than any of the measurements for the other sites, 
the darkest $B$-band
sky surface brightness at sites other than Mt.~Graham range from 22.6
to 22.84 mag\,arcsec$^{-2}$, compared to our average value of 22.53 
mag\,arcsec$^{-2}$. Our darkest run was 22.86 mag\,arcsec$^{-2}$, 
which is marginally darker by 0.02 mag\,arcsec$^{-2}$ than the darkest 
site (Mauna Kea at solar minimum). 
Since our observations were made near solar maximum, we can expect the 
Mt.~Graham site to become darker still during periods of low solar activity 
in 2006--2007. Sites other than Mt.~Graham had $V$-band sky brightness values
that varied between 21.44 and 22.29 mag\,arcsec$^{-2}$, compared to the 
Mt.~Graham average of 21.49 mag\,arcsec$^{-2}$. Our darkest run had a $V$-band 
sky surface brightness
of 21.72 mag\,arcsec$^{-2}$, which is 0.28 mag\,arcsec$^{-2}$ darker
than the brightest site (Kitt Peak (Massey \& Foltz 2000)) and 0.48 
mag\,arcsec$^{-2}$ brighter than the darkest
site (CTIO during solar minimum (Phillips 
1997\footnote{\url{http://www.ctio.noao.edu/site/pachon\_sky/}})), 
although, again, our observations were at solar maximum. There are fewer
published sky surface brightness values in $U$ and $R$, but
where we can make a comparison (La Palma (Benn \& Ellison 1998), 
ESO (Patat 2003), and Cerro Tololo (Walker 1987)), our Mt.~Graham
averages are similar, and our darkest run was 0.08 mag\,arcsec$^{-2}$
darker in $U$ than ESO and 0.19 mag\,arcsec$^{-2}$ darker in 
$R$ than La Palma.

We can compare our measured sky brightness values to the Garstang (1989) 
predicted $V$- and $B$-band sky surface brightness values for Mt.~Graham. Garstang 
calculated $V$-band sky brightness values for very clear air, during solar
minimum, and using 1980 populations for nearby towns and cities. 
This resulted in a predicted $V$-band sky brightness of 
21.94 mag\,arcsec$^{-2}$ for MGIO at the zenith, and 21.72 mag\,arcsec$^{-2}$
at a zenith distance (z) of 45\degr. This agrees well with our darkest
run which had an
average $V$-band sky surface brightness of 21.72 $\pm$ 0.04
mag\,arcsec$^{-2}$ for z $\la$ 33.6\degr. Our measured
value is slightly brighter than what would be expected from Garstang's
predictions, but this can easily be explained by an increase in population
since 1980 and the fact that our measurements were taken near solar
maximum. Garstang also predicted $B$-band sky surface brightness values of
22.93 mag\,arcsec$^{-2}$ at z=0\degr~and 22.75 mag\,arcsec$^{-2}$ at
z=45\degr, which agrees well with our
darkest run, with an average $B$-band sky surface brightness of
22.86 mag\,arcsec$^{-2}$ for z $\la$ 33.6\degr.

To determine how nearby city lights affect sky brightness,
we plot sky surface brightness vs. the azimuth (az) of our
observations in Figure 4. Data taken during nights where Moon
illumination $\ge 20\%$ were rejected from this plot.
We normalized the sky brightness of each image to the median sky 
surface brightness for the $\sec z \le 1.3$ data in each observing run, and
arbitrarily offset data-points taken during non-photometric conditions 
from those taken during photometric nights. Open circles
(non-photometric) and solid circles (photometric) represent images
taken at mid-zenith distances ($20\degr \le z < 40\degr$), while 
asterisks (non-photometric) and triangles (photometric) represent images
taken at high-zenith distances ($z \ge 40\degr$). Vertical dotted lines
mark the general direction of three cities that may contribute to 
light-pollution at the Mt. Graham site. 

Figure 4 shows that images observed toward the North during photometric
conditions tend to have darker skies than all other directions. Darker 
northern skies are seen to a lesser extent with increasing wavelength 
(0.2, 0.1, 0.06, and 0.00 mag\,arcsec$^{-2}$ darker than the median sky in 
$U$, $B$, $V$, and $R$, respectively), and not at all in the non-photometric 
data (due to the presence of cirrus). This implies that the effect may be due 
more to the large angular distance of these pointings from the zodiacal 
belt and Milky Way than to the absence of city lights in that direction. 
Phoenix and Tucson contribute somewhat to the sky brightness,  
with photometric skies between the two cities ($220\degr < az < 300\degr$)
brighter than the median sky by 0.1 mag\,arcsec$^{-2}$ in $U$, and
0.2 mag\,arcsec$^{-2}$ in $BVR$. This brightening toward Tucson and 
Phoenix is strongest at high zenith distances ($z \ge 40\degr$) and
during non-photometric conditions, which is consistent with the expected 
reflection of city lights off of clouds or cirrus.
Safford has less of an effect on
sky brightness, however, with no measurable brightening in that direction
during photometric conditions. The only exception is in the $R$-band
during non-photometric conditions, where the sky in that direction
is 0.4 mag\,arcsec$^{-2}$ brighter than the median. This might be at 
least in part due to sodium lamps from Safford, which emit at 
5500--6500 \AA, and are therefore most apparent in $R$ 
($\lambda_e \sim 6340$ \AA). This is consistent with
Massey \& Foltz (2000), who estimated the contribution of such lamps in
Tucson to be 0.17 mag\,arcsec$^{-2}$ at the zenith of Kitt Peak and 
Mt. Hopkins, with a larger effect expected at higher zenith-distances and
with the presence of clouds. Our brightest sky measurements toward Safford 
are outlying non-photometric, high-airmass data-points, 
and overall Safford contributes very little to the night sky
brightness at the location of the VATT on Mt. Graham.
Garstang (1989) predicted that the night sky would be brightest toward
Safford at a modest zenith distance of 45\degr, and considerably
brighter toward Tucson than any other direction at the extreme
zenith distance of 85\degr. However, Tucson affects our sky brightness
measurements more than Safford in almost all cases. This is in part
because the Safford lights are shielded by the mountain peak at the
VATT's location, and in part because of the strict dark-sky ordinances 
in place in Safford, as well as faster growth in Tucson than Safford 
since Garstang's 1980 population calculations. Also, smog carried up from
the Tucson valley to the nightly inversion layer likely reflects the city
lights better than the clean air above Safford. Overall, city lights have 
little affect on the sky brightness at Mt.~Graham, making it a prime 
dark-sky site.

Sky brightness can also vary with time of night, as addressed by
Walker (1988). Walker found an exponential decrease in sky brightness
at San Benito Mountain of 0.4 mag in $B$ and $V$ during the first
half of the night. Since this decrease was observed near the zenith, and was 
independent of overall sky brightness, time of year, and the presence 
of fog, Walker concluded
that it is more likely due to a natural phenomenon than a decrease
in the contribution of city lights throughout the night. Walker
mentions that this may be partially due to a decrease in the zodiacal 
light contribution throughout the night, but is likely mostly due to the
recombination of ions that were excited during the day by solar EUV 
radiation. 

We investigate this trend at Mt.~Graham in Figure 5, which shows the
dependence of sky surface brightness in $UBVR$ on fraction of the night,
where the beginning and end of the night in each run is defined as
the end and beginning of astronomical twilight for the mid-point of that run.
We plot only data-points taken during moon illumination $\le 20\%$ and 
at z $\le 40\degr$. We normalized the sky
brightness of each image to the median sky surface brightness
for the $\sec z \le 1.3$ data in each observing run, and arbitrarily 
offset data-points taken during non-photometric conditions from those 
taken during photometric nights by 1.5 mag. We approximate the nightly 
sky brightness trend with a linear least-squares fit that does not include 
measurements taken within 0.5 hours of twilight (solid lines). 
The $UB$ photometric data show no significant trend with time of night.
There is, however, a trend in photometric data in $V$ and $R$ 
(which is expected due to the nightly decrease
in \ion{O}{1} $\lambda$5577 and $\lambda$6300-34 emission line strengths), 
with a decrease in the first half of the night of 
0.1 mag\,arcsec$^{-2}$ in $V$ and 0.2 mag\,arcsec$^{-2}$ in $R$, followed 
by a slight increase in sky brightness toward the very end of the night. 
This is less than the 0.4 mag\,arcsec$^{-2}$ decrease seen in
$B$ and $V$ by Walker (1988), which may be due to the difference
in elevation of Mt.~Graham (10,400 feet) and San Benito Mountain (5248 feet).
This highlights one of the advantages of Mt. Graham's high elevation,
which contributes in many ways to making it a particularly dark site.
Non-photometric data shows a stronger trend, with an overall
decrease in sky brightness throughout the night of 0.2, 0.3, 0.3,
and 0.4 mag\,arcsec$^{-2}$ in $U$, $B$, $V$, and $R$, respectively. 
The reason for this decrease in sky brightness is uncertain at this
time, but may be related to a general decrease of cloud-cover throughout 
the night, which we often recorded in the observing logs. Local 
humidity-driven weather induced by Mt.~Graham itself may be responsible
for this, especially in late spring--early fall, when the humidity is
higher.

\section{Trends in Estimated Seeing, or Stellar FWHM at the VATT}

\subsection{Measuring the Stellar FWHM}

The FWHM of stars measured with the VATT 2kCCD is affected by
the telescope focus in addition to atmospheric effects. 
The actual focus value depends on several factors, such
as optics, temperature, airmass, and filter. Since the VATT has a fast
$\sim$ f/1 primary mirror, its focus is very sensitive to changes
in temperature during the night. Once the telescope has reached
equilibrium with the night air, the automated telescope software 
adjusts the focus to account for temperature and airmass 
changes. Particularly at the start of each night, however, it is 
necessary for the observer to frequently refocus the telescope as the 
temperature drops. Also, as the focus changes throughout the night the FWHM 
may deteriorate progressively over time, which raises the average 
stellar FWHM values with 
respect to the actual atmospheric seeing. Consistently rechecking the
focus throughout the night can minimize this effect. 
Since these data were taken as part of a galaxy survey that focuses
mainly on $U$-band galaxy surface photometry, we typically only focused
in $U$. The change in focus between filters is small, since all of the 
filters are nearly par-focal, but focusing only in $U$ may have resulted
in a slightly larger average seeing value in $B$, $V$, and $R$ than
could have been obtained if the images had been focused in each filter 
separately.  Therefore, we offer a cautionary note that the FWHM values
in our galaxy images are likely larger than what we could achieve at the 
VATT if they each
had been focused in their particular target filter, and if each galaxy
image had been preceded by a focus check. 
Also, since the FWHM's from the galaxy images presented in this paper 
were measured from stacked images, they will be marginally larger than 
if we measured them from the individual images. This is due to 
small errors in image alignment from the applied integer shifts.

We measured the stellar FWHM for all of our stacked galaxy images with 
the LMORPHO package (Odewahn, et al. 2002), which imports a list of all 
sources and their FWHM's produced with SExtractor (Bertin \& Arnouts 1996). 
Stars are 
selected from the source list for each image by interactively defining 
limits on a plot of FWHM vs. magnitude, like the one shown in Figure 6. As 
can be seen on this plot, the FWHM of stars does not significantly depend 
on their brightness (except for bright saturated stars), while brighter 
galaxies tend to be larger in size, creating a quick way of identifying stars.
This semi-automated method works well for most galaxy images, although
problems may occur for fields that contain very few bright stars. In 
such cases, our seeing estimate may be too large, since the star selection
may be contaminated by some faint extended objects. 

To obtain more accurate measurements of the atmospheric seeing than can
be measured with the galaxy images, we also
measure the FWHM of the stars with the best focus in our focus 
exposures using \texttt{IMEXAM} within
IRAF.\footnote{IRAF is distributed by the National Optical Astronomy
Observatories, which are operated by the Association of Universities
for Research in Astronomy, Inc., under cooperative agreement with the
National Science Foundation.} These focus exposures are single images in
which 5-7 short exposures at different focus settings are recorded,
where prior to each exposure the charge on the CCD is shifted by 50-100
pixels. Because these exposures are short, the stellar images are not
affected by tracking and guiding errors or by telescope vibrations (as
we will show below, this was particularly a problem in our
earlier runs). Independent FWHM measurements by two of us (Taylor and
Jansen) agree to within the measurement errors (typically 
$\sim$0.05--0.10\arcsec).

\subsection{Estimated Seeing Results}

Figure 7 shows the median FWHM of stars measured in our
stacked galaxy images as a function of airmass in $UBVR$, and is split into 
separate panels
for each observing run. There is a clear trend of increasing FWHM with
airmass, which is to be expected from the theoretical 
relation\footnote{http://www.ing.iac.es/Astronomy/development/hap/dimm.html} of
\begin{equation}
\mathrm{FWHM}(z) = \mathrm{FWHM}(0)\,sec(z)^{0.6}
\end{equation}
but, like the sky surface brightness,
this trend does not seem to have a particularly consistent slope from one
run to the next (possibly because the automatic focus did not correct
for airmass dependence accurately enough). Stacked images that are
comprised solely of individual images taken during photometric conditions
(variation in magnitude zeropoint throughout the night $\la 3\%$) are plotted 
as asterisks, while those comprised of images taken during non-photometric
conditions are plotted as open circles. This reveals that there does not seem
to be a clear trend of seeing with photometricity. However, we note that
in the two runs (April 1999 and May 2000) where there is a significant 
difference
between the seeing in the photometric nights and in the non-photometric
nights, the non-photometric nights had better seeing. The observation
log sheets noted the presence of cirrus, which is often correlated with stable
air and better seeing. Solid squares in this plot represent the FWHM of 
the stars with the best focus in the short focus exposures. 
These FWHM values tend to be smaller
than or equal to the stellar FWHM measured in galaxy images taken immediately
after the focus exposures, for the reasons mentioned in the previous
section. As the telescope focus degrades with time between focus exposures,
the stellar FWHM in the galaxy images will increase. Thus, the focus FWHM 
values are indeed a more accurate measurement of the atmospheric seeing. 

Figure 8 shows
the median low airmass ($\sec z < 1.2$) FWHM values for each run, with
solid circles representing the stellar FWHM in the stacked galaxy images,
and open circles representing the best focus FWHM in the focus exposures. 
In almost all cases, the median FWHM in the focus exposures is smaller
than that in the galaxy images, as expected. 
Except for the February 2001 observing run, which had particularly good 
seeing, it is apparent that the average FWHM values and their
uncertainties (which reflect the range of the data) are much larger
for the runs before May 2001. This
change in FWHM values corresponds to an engineering run at the
telescope in Summer 2001, during which time a vibration in the secondary
mirror mount that had contributed up to $0.4\arcsec$ to the FWHM was
removed (M. Nelson, private communication).
Adjustments were also made to the pointing map in Fall 2001. 
As Figure 8 shows, both of these improvements resulted in a significant
reduction of the FWHM of the VATT PSF. Table 2 lists the average 
of the median stellar FWHM values in the galaxy images for all runs
(ignoring the outlying April 2001 run) before and after the improvements.
There was an overall improved seeing of about $0.45\arcsec$ in all filters, 
as well as a more stable focus, as can be seen in the decreased FWHM 
scatter between these two time periods in Figure 7, and the smaller 
uncertainties in Figure 8 and Table 2. 
After the improvements, we were able to obtain sub-arcsecond seeing in
one of our combined images in $R$ in April 2002 (see Figure 7), even though
we focused in a different filter, and routinely measured sub-arcsecond seeing
in the focus frames.

The stellar FWHM values from the galaxy images are useful in determining
the average FWHM that one might realistically achieve in long (3--20
minute) object 
exposures at the VATT, with better results possible with more frequent
focusing, and with refocusing done for each filter. However, the best FWHM 
values are obtained through the shorter (several second) focus exposures. 

We can inter-compare the
FWHM in focus exposures taken in different filters by determining
the offset in the PSF between filters, which is a result of both
the wavelength dependence of atmospheric seeing 
and the contribution of the telescope. Atmospheric seeing has been
studied extensively in the past (e.g. Kolmogorov 1941, Tatarski 1961,
and Fried 1965), and has been reviewed and summarized more recently by Coulman 
(1985) and Roddier (1981). The {\it Fried parameter}, r$_{0}$, is a measure 
of the average effective size at a given wavelength, $\lambda$, of the elements 
of air
that are responsible for the angular deviations of light from a distant
point source, which is the cause of atmospheric seeing. Where
r$_{0}\propto\lambda^{6/5}$, the FWHM measured in seeing estimates is related
to r$_{0}$ by
\begin{equation} 
\mathrm{FWHM} = 0.98\,\lambda\,r_{0}^{-1} 
\end{equation}
which results in a dependence of the FWHM on wavelength of $\lambda^{-1/5}$.
To test this relation and find the FWHM contribution from the telescope, 
we plot the stellar FWHM of our images 
in each filter minus the stellar FWHM for that field in the $R$ band
(Figure 9).
We only include galaxies where exposures in each filter were taken 
immediately after one another in order to limit the effects of
airmass and large-scale seeing changes between exposures during the night. 
The outlying
points were likely due to fields imaged during highly variable
seeing conditions, or to fields where a focus exposure was taken
in between observations. The solid curve in Figure 9 traces the
theoretical $\lambda^{-1/5}$ FWHM dependence, while the crosses mark the
median FWHM$_{\lambda}$ -- FWHM$_{R}$ offset from the $\lambda^{-1/5}$ 
relation. The slight offset between the observational medians and the
theoretical $\lambda^{-1/5}$ line gives the systematic contribution
of the telescope to the wavelength dependence of the stellar FWHM.
The scatter in this plot gives a measure of the random contribution
of the atmosphere and telescope to the wavelength dependence, which can
be due both to atmospheric variations and telescope vibrations (which
is more important for the earlier runs, before the telescope improvements
made in Summer and Fall 2001). These factors cannot be separated from
one another in this plot, but we can put an upper limit on the
random contribution from the telescope as the standard deviation in
the points divided by $\sqrt 2$ (since the errors in the target filter plus
those in $R$ combine in quadrature), which is $\simeq 0.1\arcsec$ in 
all filters.

In order to more carefully determine the telescope contribution to
the wavelength dependence of the seeing, we plot the offsets between 
observation and the $\lambda^{-1/5}$ relation as a function of
FWHM as measured in $R$ in Figure 10. Points with offsets from theory greater
than 0.3\arcsec, which is significantly larger than the standard deviation of
about 0.2\arcsec, 
are rejected in order to exclude outliers caused by variable atmospheric
seeing. Visual inspection of these plots reveal that the telescope's
contribution to the wavelength dependence of stellar FWHM has no
clear dependence on FWHM, which suggests a constant offset for
all cases. The median FWHM$_{\lambda}$ -- FWHM$_{R}$ offsets from theory
found in this graph (0.006\arcsec\ for $\lambda = U$, 0.055\arcsec\ for
$\lambda = B$, and -0.050\arcsec\ for $\lambda = V$) provide a measure of 
the telescope contribution to
the FWHM wavelength dependence, which is small and well within the 
standard deviation of the observed FWHM's for all images. This telescope
contribution, plus the atmospheric contribution given by the $\lambda^{-1/5}$
relation, have been applied to the FWHM in each filter to reduce it to the 
FWHM that would have been measured in an $R$-band exposure adjacent
in time in Figure 11.

Figure 11 shows a plot of all focus FWHM values in our nine
April 1999 -- April 2002 observing runs, plus focus FWHM values for eight 
additional observing runs conducted by one of us (Jansen) for other 
projects spanning 
November 2001 -- December 2003. All values have been reduced to the
$R$-band using the $\lambda^{-1/5}$ theoretical relation, plus
the observational telescope offsets from theory found in Figure 10.
The nine April 1999 -- April 2002 runs have values that are
consistent with the eight November 2001 -- December 2003 runs, even though
each data set was observed and analyzed independently. The additional
runs give us better statistics for more recent years, and
thus verify that the observing runs before the telescope improvements 
(those to the left of the dotted
line, which marks the end of the improvements in October 2001) have
overall worse stellar FWHM values and larger scatter than those after
the telescope improvements. The observing run with the worst individual
FWHM measurements was noted to have strong winds from the northeast,
which is well-known to cause bad atmospheric seeing conditions at the VATT.
Under the best conditions, we were able to measure sub-arcsecond seeing
for many of the focus exposures, especially after the telescope improvements
in Summer and Fall 2001.

Table 2 summarizes the stellar FWHM's in the galaxy images and the
focus exposures. Median stellar FWHM values range from 0.97\arcsec\
to 2.15\arcsec\ in $R$ focus exposures, and 1.25\arcsec\ to 2.40\arcsec\
in $R$ galaxy images. The best stellar FWHM measured was 0.65\arcsec\ in
an $R$ focus exposure, and 0.95\arcsec\ in an $R$ galaxy image. This amounts 
to a linear increase in FWHM of 0.25\arcsec\ -- 0.30\arcsec\ in long exposures,
which is partially due to vibrations and variable atmospheric seeing, and 
partially due to the fact that
galaxy images may not have been taken at the best telescope focus.
Different values may be measured at other telescope sites on Mt.~Graham,
since there may be a significant telescope contribution to these values,
and trees as high as the dome surrounding the VATT site negatively impact
the seeing.

\section{Conclusions}

Figures 1 and 3 and Table 1 suggest that Mt.~Graham has a similar average
sky brightness as other dark sites, and can occasionally have 
darker skies than some of the sites reviewed here. We have found
that the sky brightness is highly variable with time, both throughout
a single observing run and from one run to the next, which is consistent with
other findings, as in Krisciunas (1997), who mentions that
except for the solar cycle, the most important effect on sky brightness
is random short term variations on timescales of tens of minutes. 
This makes it difficult to compare sky values from site to site. A more 
reliable way of comparison would be to amass a large collection of sky surface
brightness data over years at each site in order to better understand
and remove the short and long term variations in sky brightness, which is
currently not fully understood. Various site-dependent factors should also
be taken into consideration, such as the linear dependence of sky surface 
brightness on geomagnetic latitude due to Aurora effects in the Van 
Allen belt, so that low geomagnetic latitudes have somewhat darker 
skies than higher latitudes. The direction
of pointings towards cities can also affect the sky brightness, with
Tuscon and Phoenix city lights slightly increasing the sky brightness at 
the VATT in that direction by 0.1 mag\,arcsec$^{-2}$ in $U$ and 
0.2 mag\,arcsec$^{-2}$
in $BVR$. However, measurements made toward the nearest city, Safford,
are not measurably brighter than other directions (thanks to dark sky
ordinances in Safford and shielding from the mountain peak at the VATT site).
Nightly trends are also seen, with sky brightness values decreasing throughout
at least the first half of the night by an amount that depends, at least,
on the elevation of the observing site. Mt.~Graham's high elevation
contributes in this and many other ways to darker night skies, and
the minimal effect of city lights at this location make Mt.~Graham
a prime dark-sky site that can easily compete with other dark sites
around the world.

The FWHM of stars in images we took at the 
VATT have improved considerably (by $0.45\arcsec$) since 
maintenance operations for the Summer and Fall of 2001 corrected 
secondary mirror vibrations and improved the telescope pointing map. 
Figures 7, 8 and 11, and Table 2 show our stellar FWHM results. We were able 
to get sub-arcsecond seeing on occasion, especially in short (several
second) focus exposures, which are less affected than long object exposures
by vibrations, variable
atmospheric seeing, and slipping of the telescope out of focus as the
temperature changes. Because of this, the FWHM values given by focus
exposures are closer to the true atmospheric seeing (by about 0.3\arcsec) 
than those from faint object images. 

It should also be noted that there may be a significant telescope
contribution to the seeing measured at the VATT, and that
the atmospheric seeing may be different at other likely locations 
on Mt.~Graham since
the presence of trees as tall as the dome around the VATT have a negative
impact on the seeing at that telescope. 
Good seeing is not crucial to our purposes of performing surface photometry
on extended galaxies, but observers who desire smaller point spread functions 
(PSF's) should be able
to improve on our numbers by focusing more often (at least once an hour, and
more often at the beginning of the night when the temperature is more
unstable), and refocusing for each individual filter rather than using
the focus of one filter for all filters. It should also be noted that
the seeing is highly dependent on the weather, with strong northeasterly winds
contributing to much worse seeing. 

\acknowledgments

VAT was supported in part by the NASA Space Grant Graduate Fellowship at
ASU and in part by NASA grant GO-9824.1 and GO-9124.1. RAJ acknowledges 
partial support from NASA grant GO-9892.1.  We
wish to thank the staff at the VATT, especially Richard Boyle, Matthew
Nelson, and Christopher Corbally for their gracious help and support. 
We also wish to thank our many co-observers: Claudia Chiarenza, 
Thomas McGrath, Luis Echevarria, Hu Zhan, Seth Cohen, Stephen
Odewahn, Richard de Grijs, Corey Bartley, Joe Baker, Jason Mager, and
Kazuyuki Tamura.  RAJ wishes to extend particular thanks to Elizabeth
Barton, who was PI on several of the programs for which FWHM
measurements are presented in this paper.  We also thank the referee,
Wes Lockwood, who offered very knowledgeable and helpful comments.
On behalf of all MGIO
observers, we wish to thank the Tucson and Safford city councils for
passing strict low pressure sodium light ordinances, which make a
noticeable difference in the sky brightness at Mt.~Graham.

\clearpage

\figcaption[fig1.ps]{\footnotesize{Sky surface brightness in stacked
galaxy images taken at the VATT between April 1999 and April 2002 
in $U$, $B$, $V$ and $R$. Each of our observing runs is indicated 
in a separate sub-panel. Measurements obtained under non-photometric 
conditions are represented by open circles, while measurements from 
photometric nights (zeropoint variations $\lesssim3$\% throughout the 
night) are indicated by asterisks. Within a given run, the sky is 
brighter during non-photometric than photometric conditions. 
The sky surface brightness can be highly variable on monthly, nightly, and
tens of minutes time scales. Dotted lines represent the average values at Mt.
Hopkins/Kitt Peak (converted to broad-band from spectrophotometry) over
four nights in 1998 and 1999 (Massey \& Foltz 2000), just before
solar maximum (2000--2001).}}


\figcaption[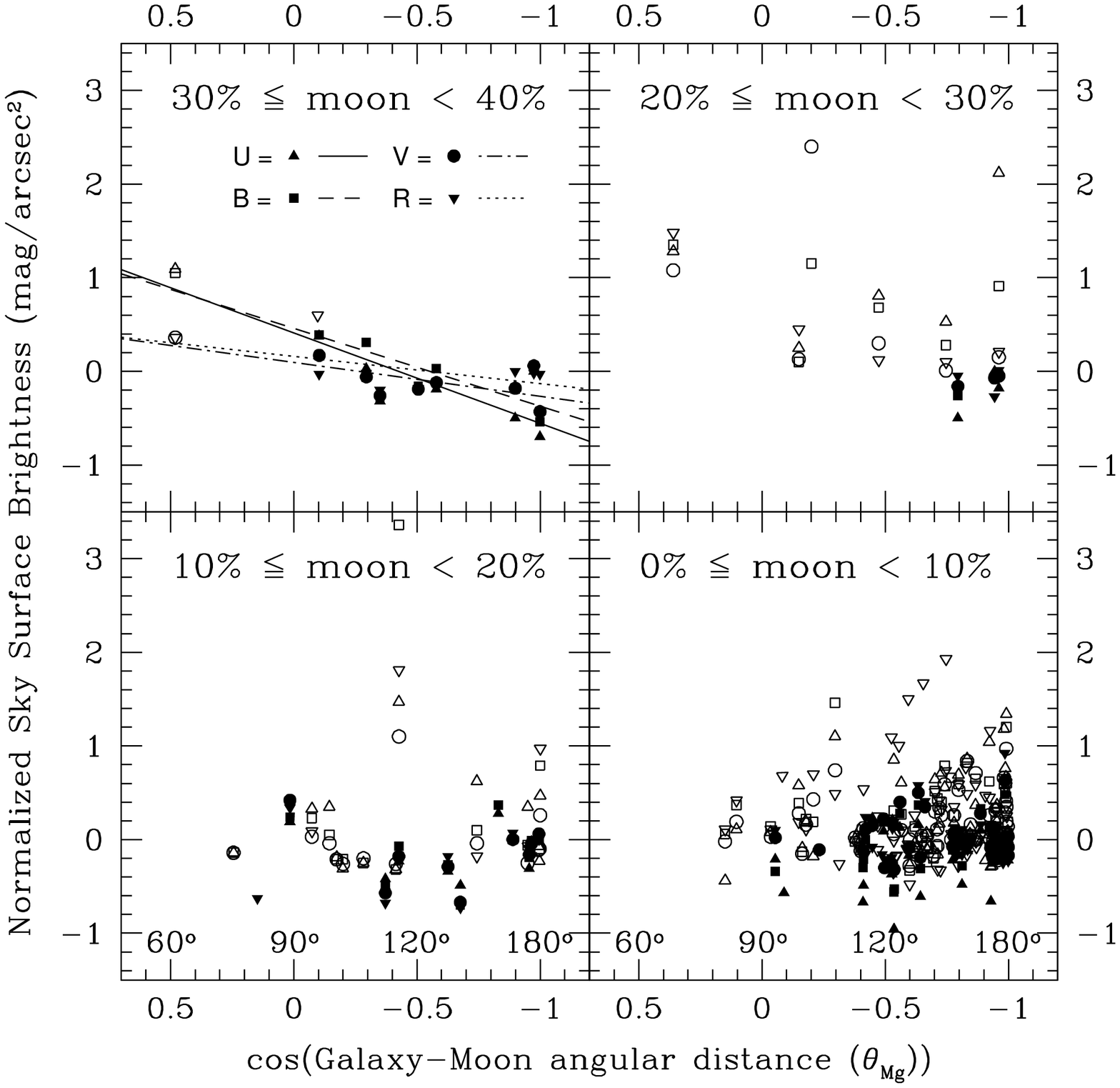]{\footnotesize{
The dependence of the sky surface brightness, normalized to the median
sky surface brightness for that observing run, on angular distance
from the Moon for different Moon illuminations. Open symbols represent
data that were taken during non-photometric nights, solid ones represent
data taken during photometric nights. Points are coded according to filter 
of observation as indicated in the upper left panel. A clear dependence on 
angular distance to the Moon is only seen for illumination $\gtrsim 20$\%.
Straight lines represent linear least-squares fits to the data in each
passband for data with $\sec z<2$.  The dependence on Moon distance is
stronger at shorter wavelengths.  In general, our galaxy images were
taken well away from the Moon and mostly during dark nights ($\lesssim
4$ days from New Moon), and thus the average sky surface brightness values
presented in this paper are not strongly affected by the Moon.
}}


\figcaption[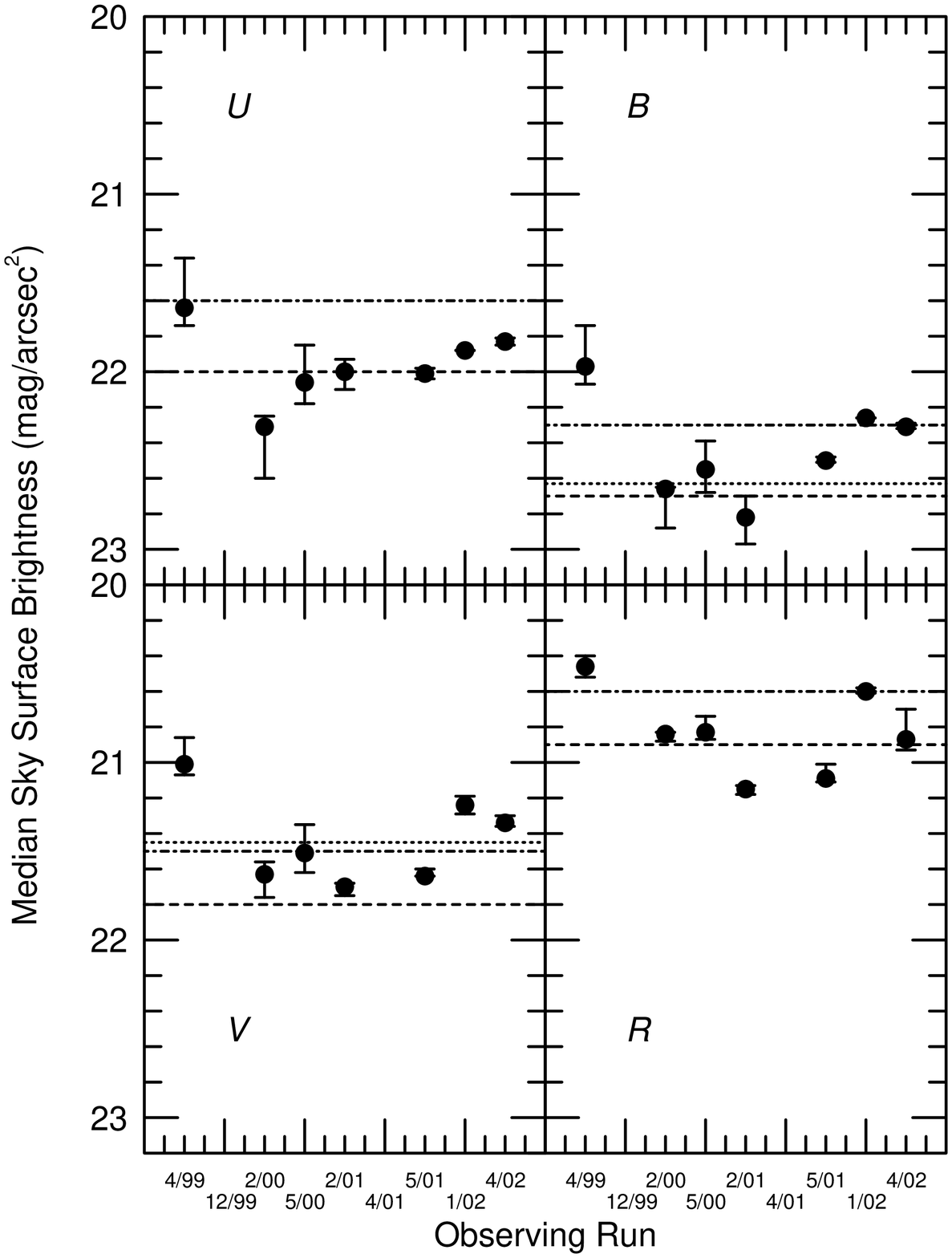]{\footnotesize{Median sky surface brightness of all
photometric stacked galaxy images taken at the VATT with airmass 
$(\sec z)  <  1.2$, rejecting no more than one obvious
outlier per data point. Error bars represent the 25\% -- 75\% quartile
range. The horizontal lines represent the average sky surface brightness
near zenith at Mt.~Hopkins and Kitt Peak, Arizona (Massey \& Foltz 2000;
converted from spectrophotometry) [\emph{dotted}], at Cerro Tololo,
Chile (Walker 1987, and 1987--1988 results at 
$http://www.ctio.noao.edu/site/pachon\_sky/$) [\emph{dashed}] 
and at La Palma, Canary Islands, Spain (Benn \& Ellison 1998) [\emph{dot-dashed}].
A comparison with other sites is given in Table 1.}}



\figcaption[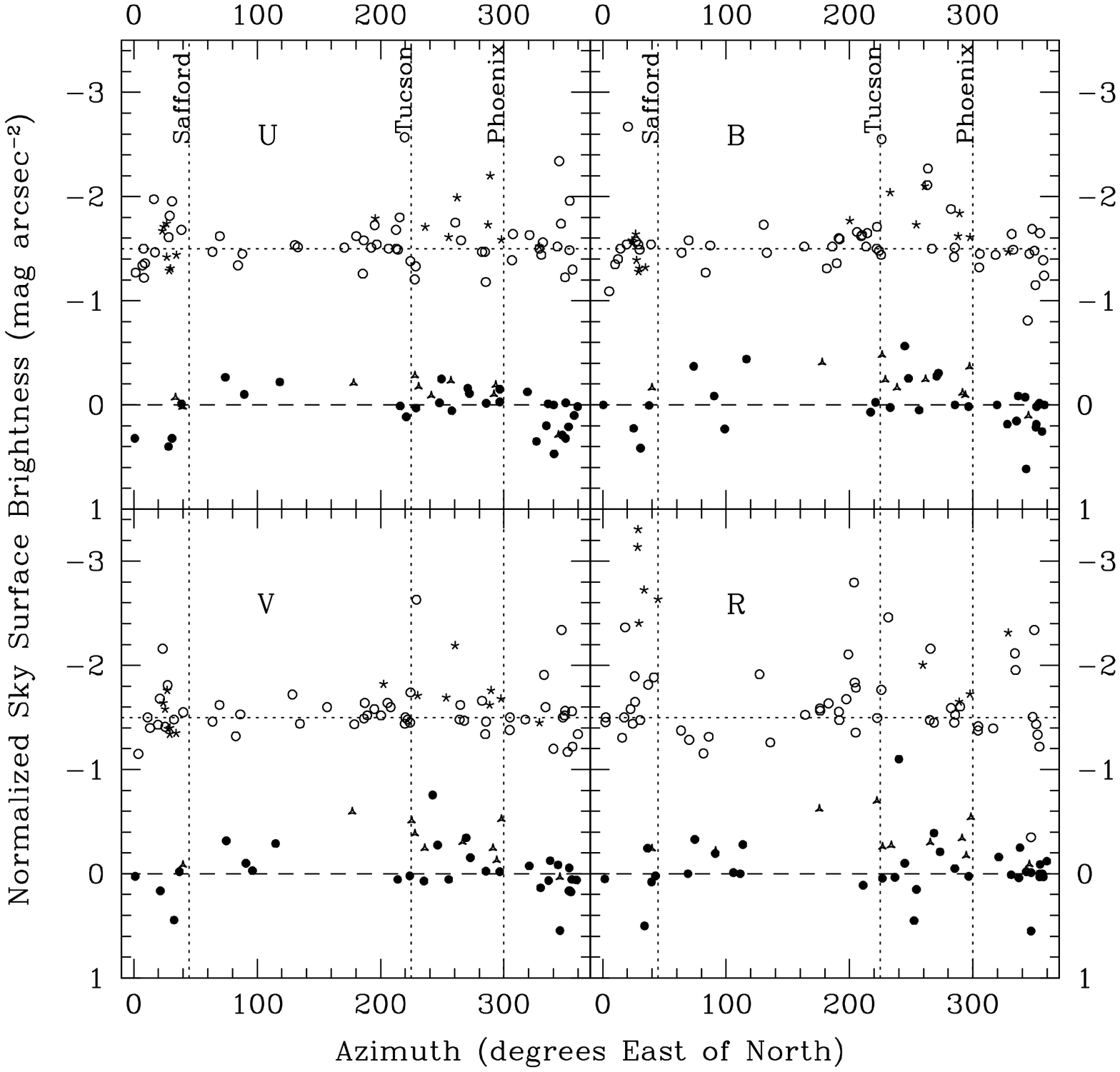]{\footnotesize{Sky brightness normalized to the
median sky for the observing run, where $\sec z \le 1.3$ and moon
illumination $\le 20\degr$. Non-photometric points (open circles for 
$20\degr \le z < 40\degr$, and asterisks for $z \ge 40\degr$)
are arbitrarily offset from photometric points (solid circles for
$20\degr \le z < 40\degr$, and triangles for $z \ge 40\degr$). 
The normalized median is marked with a dotted (non-photometric) 
or dashed (photometric) horizontal line. Vertical dotted lines mark 
the general direction of three cities that may affect the sky brightness. 
}}



\figcaption[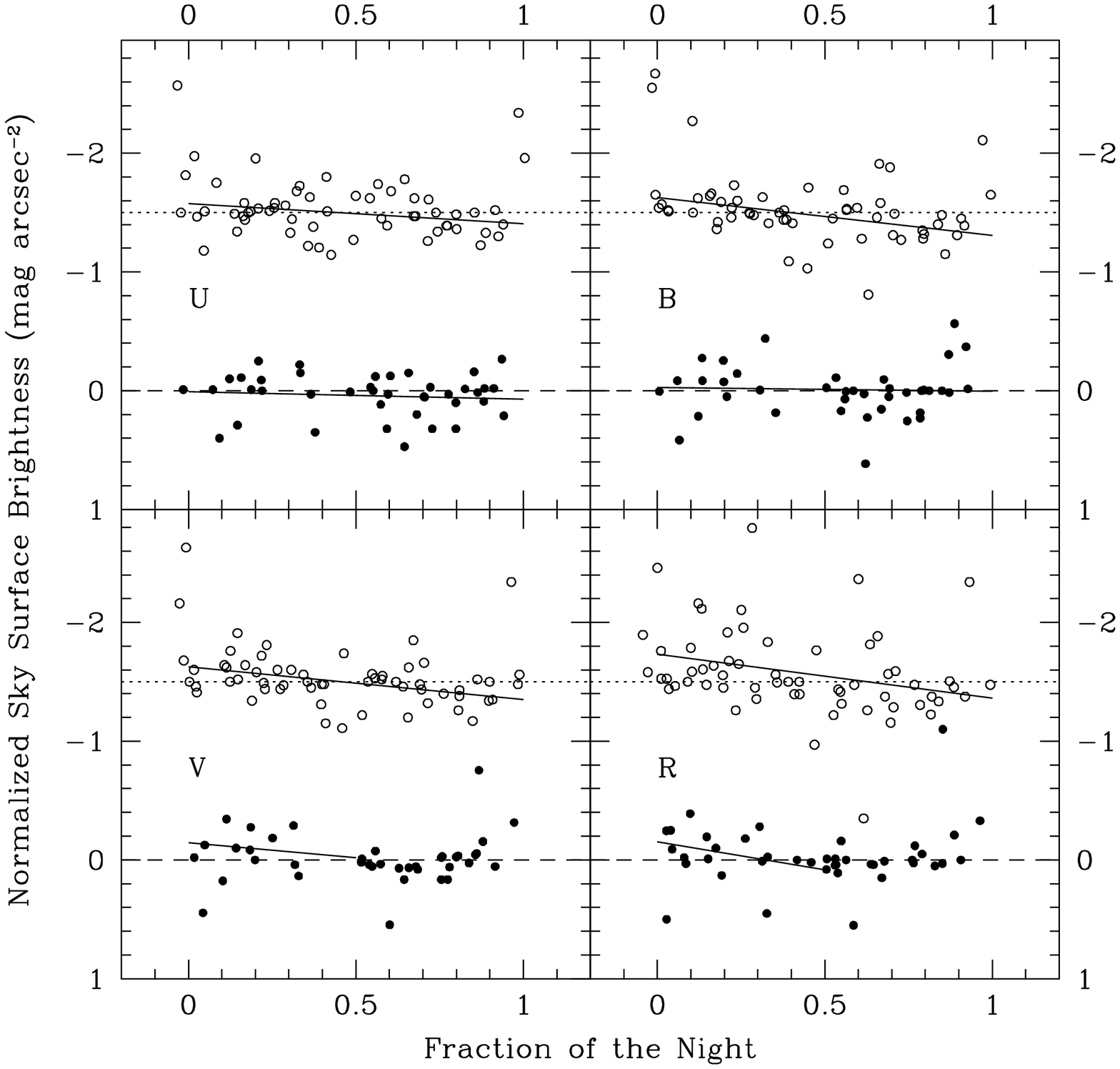]{\footnotesize{Sky brightness normalized to the
median sky for the observing run, where $\sec z \le 1.3$ and moon 
illumination $\le 20\degr$. Non-photometric points (open circles) are
arbitrarily offset from photometric points (solid circles). The normalized 
median is marked with a dotted (non-photometric) or dashed (photometric) 
line. The beginning and end of the night is defined by the end and beginning
of astronomical twilight for the mid-point of the observing run, such
that dusk is at fraction=0, and dawn at fraction=1. Solid
lines are the linear least-squares fit to the data, excluding measurements
taken within 0.5 hours of twilight.}}


\figcaption[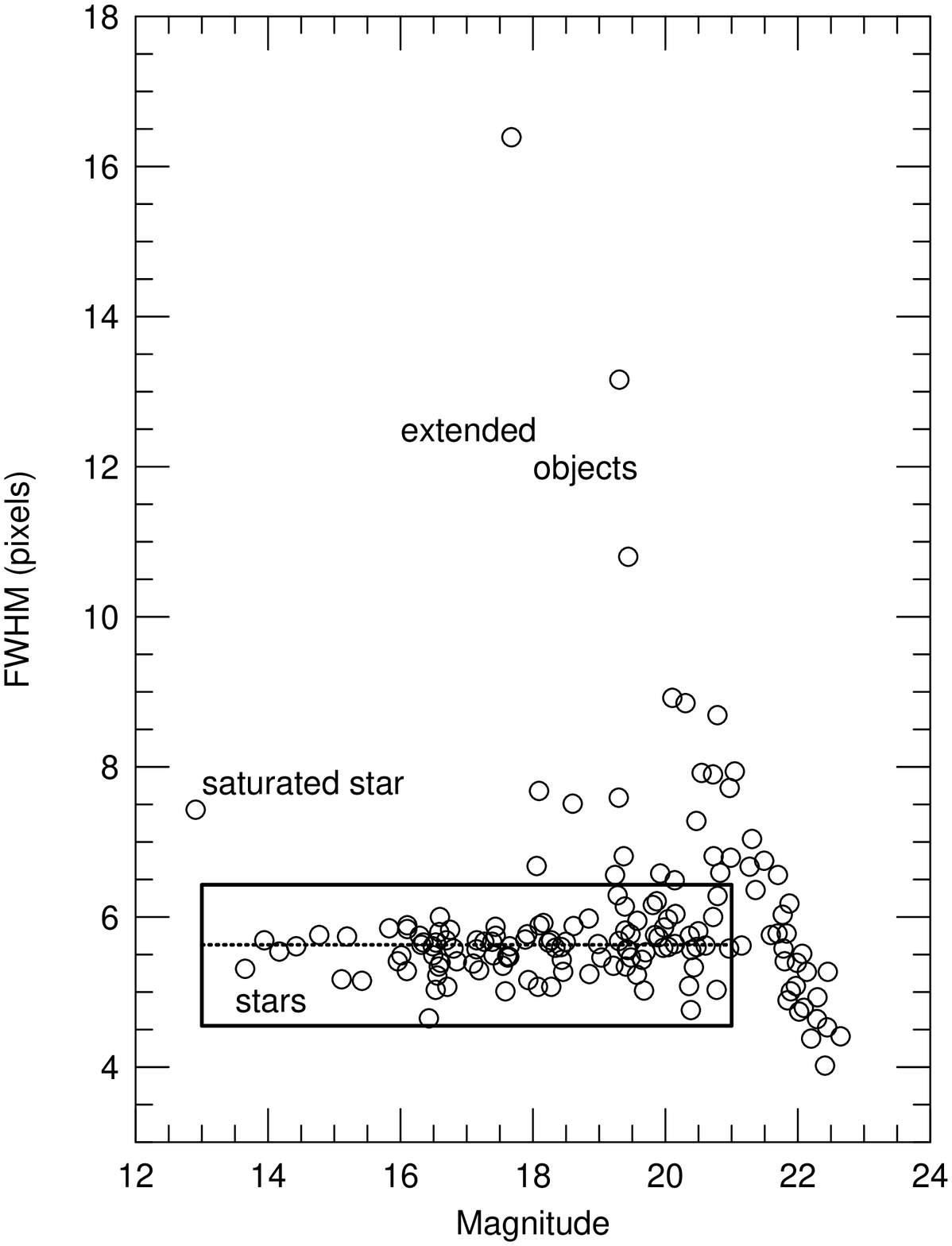]{\footnotesize{Object FWHM versus apparent magnitude in a single 
galaxy field. We use the fact that the FWHM of a star does not depend
on its brightness to separate stars and extended objects, as labeled on
the plot. For the purpose of our semi-automated seeing measurements, 
we excluded saturated stars and stars that are too faint to yield reliable 
measurements. The solid box encloses the objects that were used to compute 
the mean stellar FWHM for this field (\emph{dotted horizontal line}).}}


\figcaption[fig7.ps]{\footnotesize{Median stellar FWHM in images 
taken at the VATT between April 1999 and April 2002 in $U$, $B$, $V$ 
and $R$. Each of our observing runs is indicated in a separate sub-panel.
Measurements obtained under non-photometric conditions are represented
by open circles, while measurements from photometric nights (zeropoint 
variations $\lesssim3$\% throughout the night) are
indicated by asterisks. The solid squares represent the stellar FWHM
corresponding to the best focus setting as measured in short focus
exposures. These tend to be smaller than or equal to the FWHM's measured in
adjacent object exposures. We typically focused the telescope in $U$, since
that is where most of our galaxy images would be taken.
}}


\figcaption[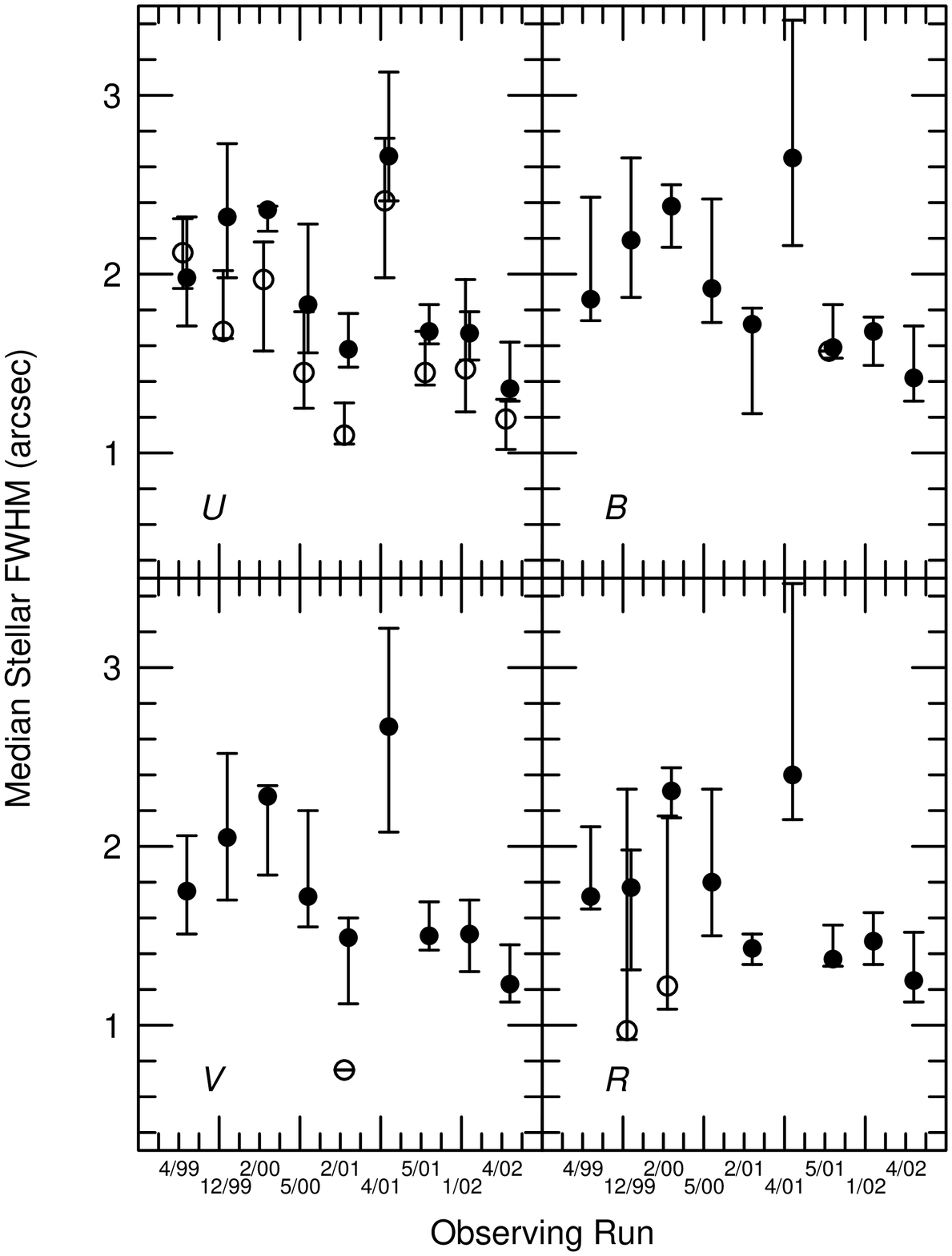]{\footnotesize{Historical trend in our FWHM measurements. 
{\bf Solid circles:} Median stellar FWHM at low airmass ($\sec z < 1.2$) 
measured in our stacked galaxy images.
{\bf Open circles:} Median FWHM of the best focus setting measured in 
short focus exposures. Error bars represent the 25\% --
75\% quartile range for each run. Improvements to the telescope in Summer 
and Fall 2001 significantly reduced the stellar FWHM's measured during the 
later runs.
}}


\figcaption[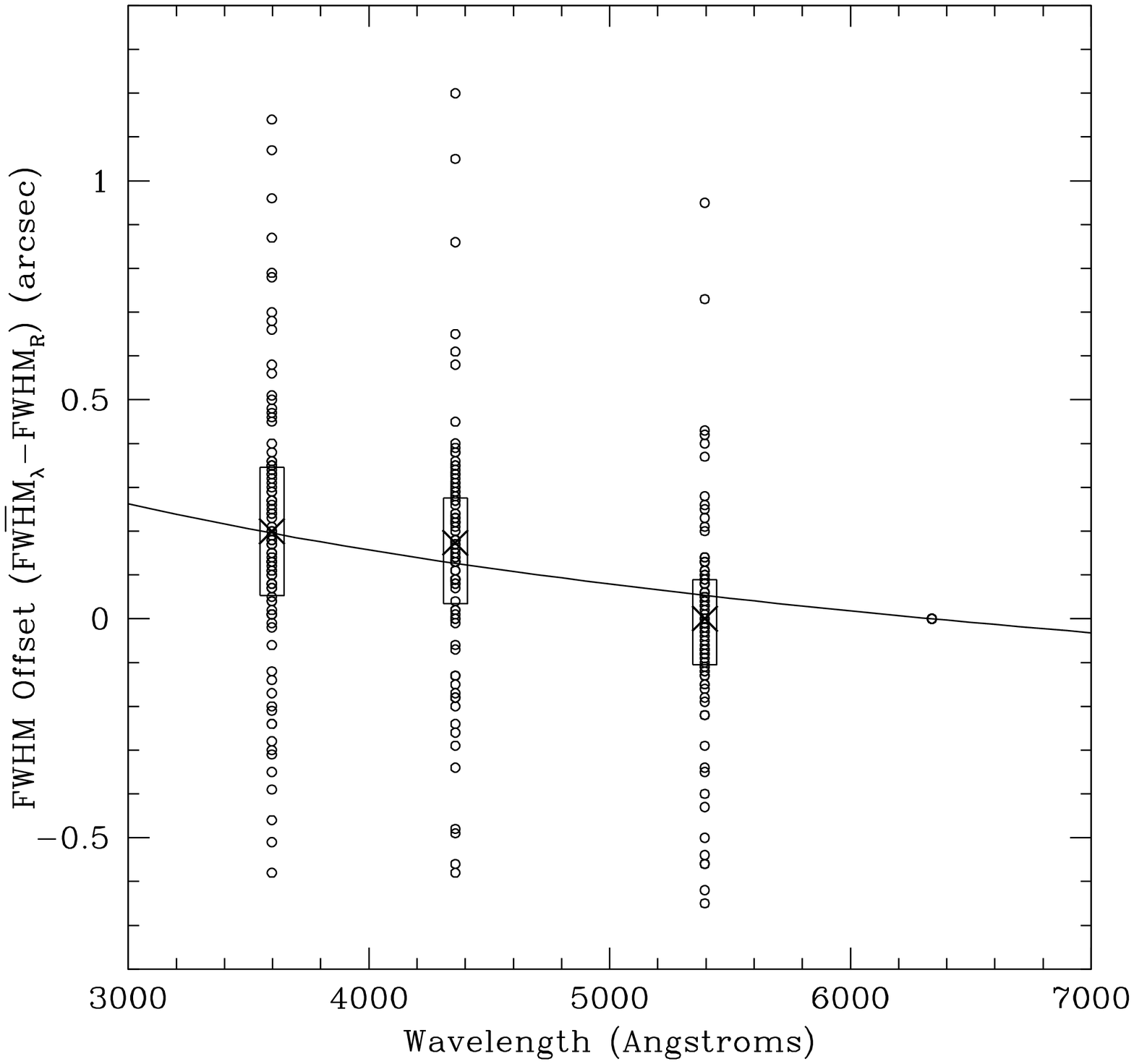]{\footnotesize{
{\bf Open circles:} stellar FWHM of each
stacked galaxy image minus the stellar FWHM in the $R$-band image
of that galaxy, producing a measure of the FWHM offset between
filters at the VATT. Outliers are due to highly variable seeing
conditions or cases where a focus exposure was taken in between
observations for a single galaxy. Galaxies where observations in
each filter were not carried out immediately after one another are not
included on this plot. {\bf Crosses:} the median FWHM$_{\lambda}$ --
FWHM$_{R}$ offset from theory for each filter. The boxes surrounding the
medians enclose the 25\% -- 75\% quartile range. {\bf Solid line:} The value
offsets would have if the FWHM's followed the theoretical $\lambda ^{-1/5}$
dependence, using the median FWHM$_{R}$ value of 1.63\arcsec. The
divergence of theory from the median observed offsets are due 
to specific telescope properties at the VATT that cause a systematic
contribution from the telescope to the wavelength dependence of the seeing. 
The scatter in this plot gives random offsets from theory which are partially 
due to atmospheric variations, and partly due to telescope vibrations (which is 
particularly important for the earlier runs). This vibrational component
cannot be separated from the atmospheric effects, but it cannot be larger than 
the standard deviation of the points, which is
$\simeq 0.2\arcsec$ in all filters.}}


\figcaption[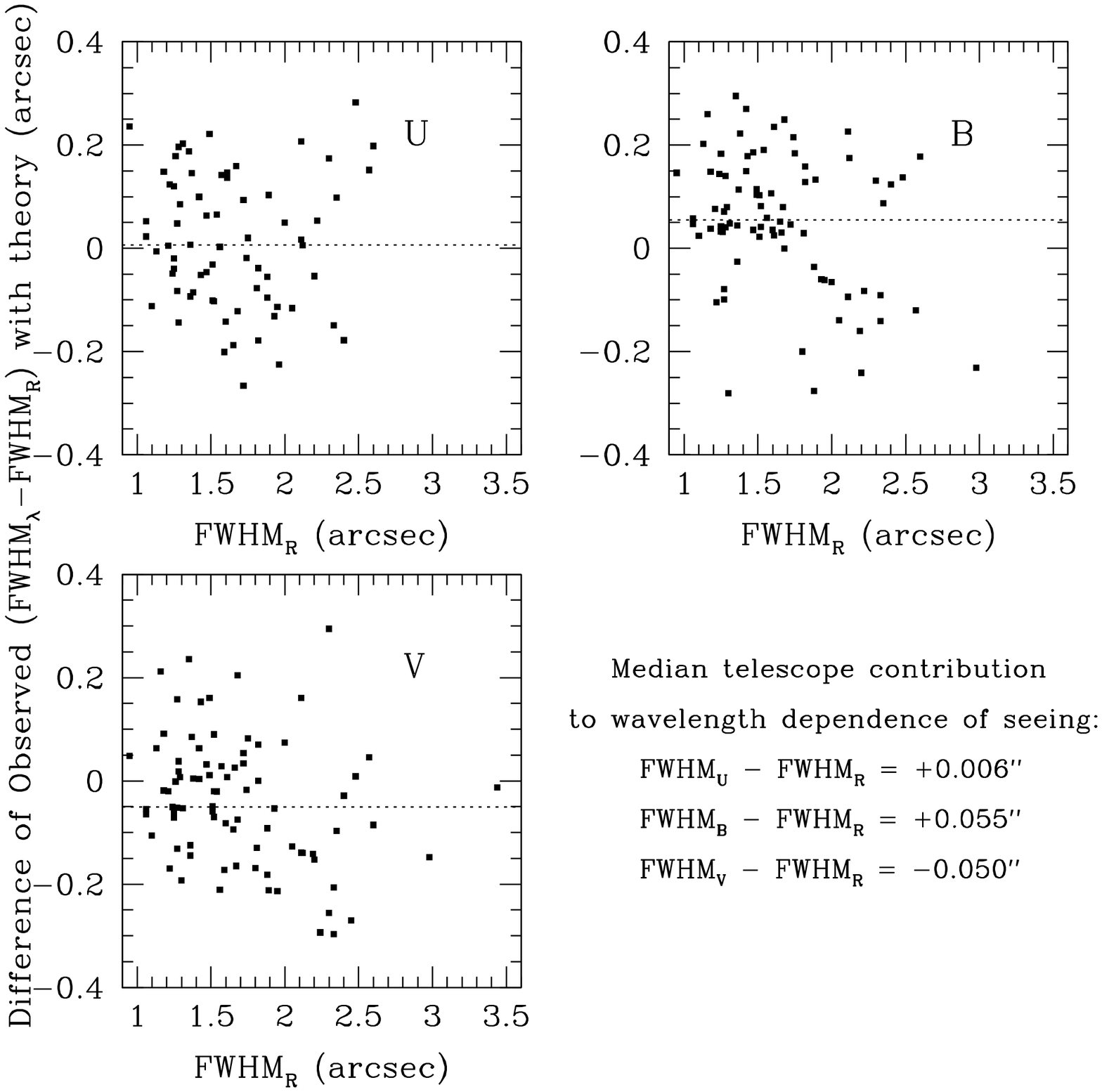]{\footnotesize{
Comparison of the theoretical $\lambda^{-1/5}$ wavelength dependence of
stellar FWHM to the observed wavelength dependence for each image, for
the purpose of reducing FWHM values to the $R$-band in order to inter-compare
focus exposures taken in different filters (as in Figure 11).
We find the observed FWHM in each passband ($UBV$) 
minus the FWHM in the reference filter, $R$, and subtract this
from the theoretical result, then plot this offset versus the observed 
FWHM in $R$. Galaxies where observations in each filter were not carried 
out immediately after one another are not included.
Points with offsets from theory greater than 0.3\arcsec\ (which is 
outside the standard deviation of 0.2\arcsec\ for all of the points) were 
rejected to avoid outliers caused by variable atmospheric conditions. 
There is no strong dependence on FWHM$_{R}\,$for this offset in
any filter, and thus we apply a constant small telescope correction to all 
FWHM values in Figure 11 of the median (observation-theory) offset in $UBV$
(listed in the figure and marked by dotted lines),
plus the atmospheric contribution given by the $\lambda^{-1/5}$
relation.}}


\figcaption[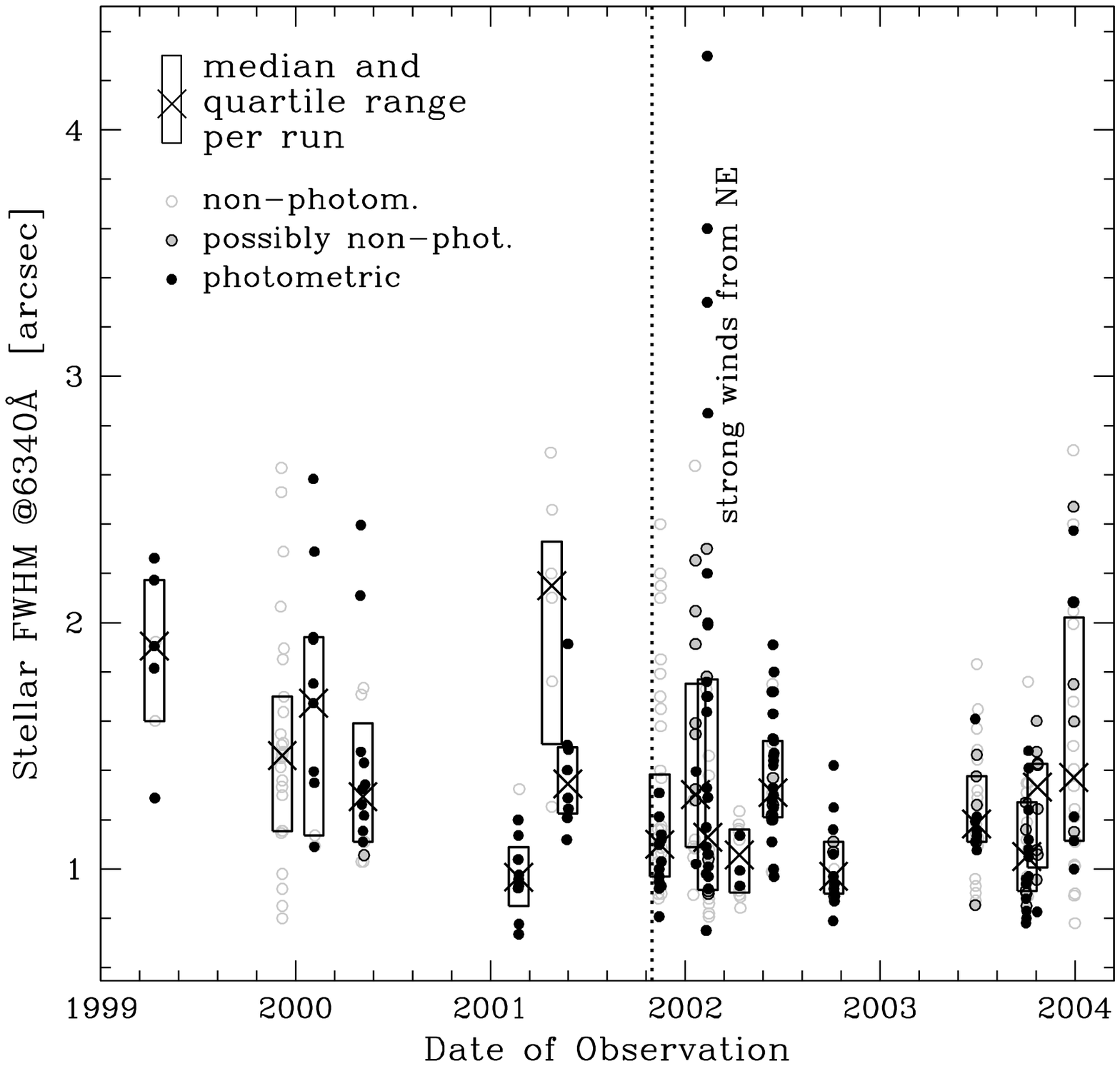]{\footnotesize{VATT focus exposure stellar FWHM values normalized
to the $R$-band using the theoretical atmospheric $\lambda^{-1/5}$ dependence
plus the observational median telescope contribution offsets found in Figure 10.
This plot includes observing runs carried out    
by one of us (R. Jansen), in addition to the observing runs the rest of this 
paper focuses on. The worst
FWHM values were measured when strong winds were blowing from the
Northeast, which always results in particularly bad seeing conditions
at the VATT. Sub-arcsecond $R$-band seeing was reached on occasion 
throughout this time period. FWHM values are highly variable, although
an overall improvement in median FWHM and scatter is apparent in the 
observing runs following telescope improvements made in Summer and Fall 2001. 
The dotted line marks the end of the implementation of these improvements.
}}


\begin{deluxetable}{lllllll}
\tabletypesize{\scriptsize}
\tablecaption{Average Photometric Sky Surface Brightness ($\mu$) Near Zenith
at Various Sites. \label{Table-1}}
\tablewidth{0pt}
\tablehead{
\colhead{Site} & \colhead{Condition} & \colhead{Obs. Dates} &
\colhead{$\mu_U$} & \colhead{$\mu_B$} &
\colhead{$\mu_V$} & \colhead{$\mu_R$}
}
\startdata
Mt.~Graham\tablenotemark{1} & Darkest run & $U$: 02/00, $BVR$: 02/01 & 22.38 & 22.86 & 21.72 &
21.19\\
Mt.~Graham\tablenotemark{1} & All runs & 04/99--04/02 & 22.00 & 22.53 & 21.49 & 20.88\\
Mt.~Graham\tablenotemark{1} & Brightest run & 04/99 & 21.68 & 22.01 & 21.04 & 20.46\\
\\
Mt.~Hopkins\tablenotemark{2} & \nodata & 11/98 & \nodata & 22.63 & 21.46 & \nodata\\
Kitt Peak\tablenotemark{2} & \nodata & 11/99 & \nodata & 22.63 & 21.44 & \nodata\\
Mauna Kea\tablenotemark{3} & Solar min. & 96 & \nodata & 22.84 & 21.91 & \nodata\\
Mauna Kea\tablenotemark{3} & Solar max. & 92 & \nodata & 22.22 & 21.29 & \nodata\\
La Palma\tablenotemark{4} & \nodata & 87-96 & 22.0~ & 22.7~ & 21.9~ & 21.0~\\
ESO/La Silla\tablenotemark{5} & \nodata & 04/00--09/01 & 22.3~ & 22.6~ & 21.6~ & 20.9~\\
Cerro Tololo\tablenotemark{6} & \nodata & 87--88 & 22.0~ & 22.7~ & 21.8~ & 20.9~\\
Cerro Tololo\tablenotemark{7} & \nodata & 97 & \nodata & 22.8~ & 22.2~ & \nodata\\\enddata
\tablecomments{All sky surface brightness values have units of mag\,arcsec$^{-2}$.}
\tablenotetext{1}{Mean error on mean Mt.~Graham values $\la 0.04\,$
mag\,arcsec$^{-2}$.}
\tablenotetext{2}{Massey \& Foltz 2000. Calculated from spectrophotometry.}
\tablenotetext{3}{Krisciunas 1997.}
\tablenotetext{4}{Benn \& Ellison 1998. Solar min., high galactic and ecliptic latitude.
Measured 0.4 mag\,arcsec$^{-2}$ brighter at solar max.}
\tablenotetext{5}{Patat 2003. Values corrected to zenith.}
\tablenotetext{6}{Walker 1987, and 1987--1988 results at $http://www.ctio.noao.edu/site/pachon\_sky/$}
\tablenotetext{7}{Phillips 1997 results at $http://www.ctio.noao.edu/site/pachon\_sky/$}
\end{deluxetable}

\clearpage

\begin{deluxetable}{llllll}
\tabletypesize{\scriptsize}
\tablecaption{Median stellar FWHM measurements at the VATT. \label{Table-2}}
\tablewidth{0pt}
\tablehead{
\colhead{type} & \colhead{date} & \colhead{$U$-band} & \colhead{$B$-band} 
& \colhead{$V$-band} &
\colhead{$R$-band} 
}
\startdata
Best median focus\tablenotemark{1} & 2/01 \& 10/02 & \nodata & \nodata & \nodata
 & $0.97\arcsec \pm 0.06\arcsec$ \\
Worst median focus\tablenotemark{1} & 4/01 & \nodata & \nodata & \nodata & $2.15
\arcsec \pm 0.42\arcsec$\\
Best FWHM in single focus exposure\tablenotemark{1} & 2/01 & \nodata & \nodata &
 \nodata & 0.65\arcsec \\
\\
All galaxy images\tablenotemark{2} & 4/99--2/01 & $2.01\arcsec \pm 0.25\arcsec$
& $2.01\arcsec \pm 0.34\arcsec$ & $1.86\arcsec \pm 0.39\arcsec$ &
$1.81\arcsec \pm 0.24\arcsec$\\
All galaxy images\tablenotemark{3} & 5/01--4/02 & $1.57\arcsec \pm 0.10\arcsec$
& $1.56\arcsec \pm 0.12\arcsec$ & $1.41\arcsec \pm 0.12\arcsec$ & 
$1.36\arcsec \pm 0.08\arcsec$\\
Best median in galaxy images\tablenotemark{4} & 4/02 & $1.36\arcsec \pm 0.03\arcsec$ & $1.42\arcsec \pm 0.06\arcsec$ & $1.23\arcsec \pm 0.04\arcsec$ & $1.25\arcsec \pm 0.05\arcsec$ \\
Worst median in galaxy images\tablenotemark{4} & 4/01 & $2.66\arcsec \pm 0.12\arcsec$ & $2.65\arcsec \pm 0.22\arcsec$ & $2.67\arcsec \pm 0.26\arcsec$ & $2.40\arcsec \pm 0.11\arcsec$ \\
Best FWHM in single galaxy image & $UBV$: 2/01, $VR$: 4/02  & 1.12\arcsec & 1.12\arcsec & 1.03\arcsec & 0.95\arcsec \\
\enddata
\tablecomments{Stellar FWHM values measured in focus frames are closer to 
the true atmospheric seeing than stellar FWHM values measured in galaxy 
images, because focus exposures
are short (a few seconds compared to a few minutes) and record 
the best telescope focus (which may have deteriorated in galaxy
exposures). 
Focusing must be done frequently (at least once an hour, possibly more
at the beginning of the night and less toward the end of the night) in
order to obtain the best stellar FWHM values in deep object exposures.
For our galaxy images we typically focused in $U$. Focusing in each
filter separately would result in smaller stellar FWHM's in the other
pass-bands.}
\tablenotetext{1} {Exposures taken in filters other than $R$ were reduced
to $R$ using the theoretical $\lambda^{-1/5}$ dependence and the observed
contribution from the telescope added in quadrature. Median values are per 
observing run.}
\tablenotetext{2} {Before telescope improvements in Summer and Fall 2001.}
\tablenotetext{3} {After telescope improvements in Summer and Fall 2001.}
\tablenotetext{4} {Median values are per observing run.}
\end{deluxetable}


\begin{thebibliography}{}
\bibitem[Benn(1998)]{ben98} Benn, C.R., and Ellison, S.L. 1998, NewAR,
42, 503
\bibitem[Bertin(1996)]{ber96} Bertin, E., and Arnouts, S. 1996, AJ, 117,393
\bibitem[Coulman(1985)]{col85} Coulman, C.E. 1985, ARA\&A, 23, 19
\bibitem[Fried(1966)]{fri66} Fried, D.L. 1965, OSAJ, 55, 1427
\bibitem[Garstang(1989)]{gar89} Garstang, R.H. 1989, PASP, 101, 306
\bibitem[Kolmogorov(1941)]{kol41} Kolmogorov, A.N. 1941, in Tikhomirov, V.M., ed, Selected works of A.N. Komogorov, Mathematics and its applications (Soviet series), Klewer Academic press (1991)
\bibitem[Krisciunas(1991)]{kris91} Krisciunas, K., and Schaefer, B. 1991, PASP, 103, 1033
\bibitem[Krisciunas(1997)]{kris97} Krisciunas, K. 1997, PASP, 109, 1181
\bibitem[Landolt(1992)]{lan92} Landolt, A.U. 1992, AJ, 104, 340
\bibitem[Massey(2000)]{mas00} Massey, P., and Foltz, C.B. 2000,
PASP, 112, 566
\bibitem[Odewahn(2002)]{ode02} Odewahn, S.C., Cohen, S.H., Windhorst, R.A.,
Philip, N.S. 2002, 568, 539
\bibitem[Patat(2003)]{pat03} Patat, F. 2003, A\&A, 400, 1183
\bibitem[Roddier(1981)]{rod81} Roddier, F. 1981, in Wolf E., ed, Progress in Optics 19, North Holland Publ., Amsterdam 
\bibitem[Tatarski(1961)]{tat61} Tatarski, V.I. 1961, Wavefront Propagation in a Turbulent Medium, Dover, New York 
\bibitem[Walker(1987)]{wal87} Walker, A. 1987, NOAO Newsletter, No. 10, 16
\bibitem[Walker(1988)]{wal88} Walker, M. 1988, PASP, 100, 496
\end{thebibliography}
\end{document}